# Electrochemical synthesis and properties of $CoO_2$, the $x = 0$ phase of the $A_xCoO_2$ systems ($A$ = Li, Na)

T. Motohashi[1], T. Ono[1,2], Y. Katsumata[1,2], R. Kanno[2], M. Karppinen[1,3], and H. Yamauchi[1,2,3]

[1]*Materials and Structures Laboratory, Tokyo Institute of Technology, Yokohama 226-8503, Japan*

[2]*Interdisciplinary Graduate School of Science and Engineering, Tokyo Institute of Technology, Yokohama 226-8502, Japan*

[3]*Laboratory of Inorganic and Analytical Chemistry, Helsinki University of Technology, P.O. Box 6100, FI-02015 TKK, Finland*

(Dated: September 5, 2007. Revised dated: September 12, 2007)

Single-phase bulk samples of the "exotic" $CoO_2$, the $x = 0$ phase of the $A_xCoO_2$ systems ($A$ = Li, Na), were successfully synthesized through electrochemical de-intercalation of Li from pristine $LiCoO_2$ samples. The samples of pure $CoO_2$ were found to be essentially oxygen stoichiometric and possess a hexagonal structure consisting of stacked triangular-lattice $CoO_2$ layers only. The magnetism of $CoO_2$ is featured with a temperature-independent susceptibility of the magnitude of $10^{-3}$ emu/mol Oe, being essentially identical to that of a Li-doped phase, $Li_{0.12}CoO_2$. It is most likely that the $CoO_2$ phase is a Pauli-paramagnetic metal with itinerant electrons.

The layered cobalt oxide systems, $A_x\text{CoO}_2$ ($A$ = Li, Na), have attracted increased attention due to recent discoveries of various unconventional transport and magnetic properties. Namely, among the members of the systems, $\text{Na}_x\text{CoO}_2$ exhibits a superb thermoelectric property about $x = 0.7$ [1,2], a spin-density-wave excitation at $x = 0.75$ [3], a charge ordering at $x = 0.50$ [4], and superconductivity about $x = 0.35$ with a hydrated form [5]. The crystals of $A_x\text{CoO}_2$ consist of alternate stacking of a $\text{CoO}_2$ layer and a single atomic layer of $A$ ions. The $\text{CoO}_2$ layer contains a 2D triangular cobalt lattice with possible magnetic frustrations (due to the geometry) that may be underling aforementioned unconventional electronic states.

As mentioned above, the physical properties sensitively depend on the concentration of $A$ ions. It is thus important to establish the electronic phase diagrams for the $A_x\text{CoO}_2$ systems with respect to $x$. Nevertheless, properties have remained unclear in the low $x$ regime, i.e. $x < 0.25$, due to the difficulty in sample syntheses. In particular, it is worth synthesizing the $x = 0$ phase of the $A_x\text{CoO}_2$ systems, *i.e.* pure $\text{CoO}_2$, since it can be regarded as the parent of these systems. Synthesis of $\text{CoO}_2$ samples was previously reported [6,7], but the sample purity was insufficient for the accurate determination of physical properties. Here we report the synthesis and the properties of pure $\text{CoO}_2$ bulk samples.

Bulk samples of pure $\text{CoO}_2$ phase were synthesized through electrochemical de-intercalation of Li from pristine $\text{LiCoO}_2$ samples. The $\text{LiCoO}_2$ samples were prepared by a conventional solid-state reaction technique. A mixture of $\text{Li}_2\text{CO}_3$ and $\text{Co}_3\text{O}_4$ with the ratio of Li/Co = 1 : 1 was calcined (600°C) and then sintered (900°C) in flowing $\text{O}_2$ gas. Electrochemical oxidation was carried out with a constant current (*i.e.* galvanometric) setup utilizing an airtight flat cell filled with a nonaqueous electrolyte. The electrochemical cell consisted of an as-synthesized $\text{LiCoO}_2$ pellet and an aluminum metal disk as the cathode and anode, respectively. No auxiliary agents (*e.g.* acetylene black and Teflon powder) were added to the bulk pellet to avoid any magnetic noise sources. Since high-valent cobalt oxides tend to experience chemical instability when exposed to atmospheric moisture, sample handling and characterization were carefully made in an inert gas atmosphere. Details of experimental procedures are given elsewhere [8].

Single-phase $CoO_2$ samples were successfully obtained through the aforementioned electrochemical synthesis technique. Figure 1 shows a typical voltage versus $x$ plot for the $Li_xCoO_2$/Al cell. The $x$ values were estimated through theoretical calculations based on Faraday's law with an assumption that the full amount of electricity was used for the Li de-intercalation reaction. The cell voltage gradually increased with decreasing $x$ in $Li_xCoO_2$ and finally reached +4.78 V at $x = 0.0$. As $x$ decreased below $\approx 0.1$, the cell voltage increased rapidly to indicate the completion of lithium extraction [6]. The actual Li content ($x$) of the resultant $CoO_2$ sample was determined by means of ICP-AES to be below the detection limit of the apparatus, *i.e.* smaller than 0.01. This ensures that our $CoO_2$ samples are indeed the $x = 0$ end member of the $Li_xCoO_2$ system.

In Fig. 2, x-ray powder diffraction (XRPD) patterns for the pristine $LiCoO_2$ and the $CoO_2$ samples are shown. The pattern for $LiCoO_2$ was readily refined based on space group *R*-3*m* with the lattice parameters $a = 2.814$ Å and $c = 14.05$ Å, being in good agreement with those previously reported [6,9]. For $CoO_2$, on the other hand, diffraction peaks were indexed based on space group *P*-3*m*1 with the lattice parameters $a = 2.820$ Å and $c = 4.238$ Å. The $LiCoO_2$ phase crystallizes in a so-called O3-type structure, in which Li ions occupy the octahedral site with three $CoO_2$ layers per unit cell, while the $CoO_2$ phase possesses an O1-type structure consisting of a single $CoO_2$ layer only per unit cell. Oxygen-content analysis was performed based on hydrogen reduction experiments in a 5% $H_2$/Ar gas flow with a thermobalance. The oxygen content of the sample was determined to be $CoO_{1.98\pm0.02}$. Thus, the resultant $CoO_2$ sample is essentially oxygen stoichiometric, implying that the highest formal valence of cobalt is realized, *i.e.* $V_{Co} \approx +4.0$.

In the cell voltage versus $x$ curve shown in Fig. 1, one may notice that there appears a narrow plateau at about $x = 0.12$ prior to the rapid increase in voltage, suggesting the existence of another stable phase neighboring on the $x = 0$ end member. We thus synthesized a $Li_xCoO_2$ sample precisely controlling the electric charge setting at $x = 0.12$. It is found that the $x = 0.12$ sample is indeed of single phase possessing a hexagonal structure with the lattice parameters $a = 2.821$ Å and $c = 27.13$ Å (Fig. 2). The crystal of the $x = 0.12$ phase is reported to consist of alternate stacking of a Li-intercalated O3-type block (as in $LiCoO_2$) and a Li-free O1-type block (as in $CoO_2$), leading to a six-$CoO_2$-layer unit cell that is called "H1-3" [10,11]. Note that $Li_{0.12}CoO_2$ ($V_{Co} = +3.88$) may correspond to an electron-doped phase of $CoO_2$ ($V_{Co} = +4$).

Magnetic susceptibility ($\chi$) measurements were performed with a SQUID magnetometer (MPMS-XL; Quantum Design). The $\chi$ versus $T$ plots for the $LiCoO_2$ (pristine), $Li_{0.12}CoO_2$, and $CoO_2$ samples are given in Fig. 3. The $\chi$ value for the $LiCoO_2$ sample is small in magnitude and little dependent on temperature, due to a nonmagnetic nature of low-spin $Co^{III}$. Surprisingly, the $CoO_2$ and $Li_{0.12}CoO_2$ samples also exhibit temperature-independent susceptibility in a temperature range between 50 and 300 K. The susceptibility data for $CoO_2$ are in good agreement with those reported by C. de Vaulx *et al.* [7], although their sample contains a large amount of secondary phases. The $\chi$ versus $T$ plots were fitted with the following formula: $\chi = \chi_0 + C/(T - \Theta)$, where $\chi_0$, $C$, and $\Theta$ denote a constant susceptibility, the Curie constant, and the Weiss temperature, respectively. Fitted values for $\chi_0$, $C$, and $\Theta$ are summarized in Table 1. The $C$ value for the $CoO_2$ sample yields an effective magnetic moment ($\mu_{eff}$) of 0.18 $\mu_B$/Co site, being much smaller than the theoretical spin-only value of low-spin $Co^{IV}$ ($S = 1/2$), *i.e.* 1.73 $\mu_B$/Co site. Thus, any localized spin model is unlikely.

We conclude: (1) the magnetism of $CoO_2$ is featured with a temperature-independent susceptibility with a relatively large value for $\chi_0$, and (2) the upturn behavior at low temperatures is likely originating from an extrinsic cause, *e.g.* lattice defects. A possible (and the most conservative) explanation for this is that the $CoO_2$ phase is a Pauli-paramagnetic metal with itinerant electrons. Assuming that the difference in magnitude of $\chi_0$ between $CoO_2$ and $LiCoO_2$ corresponds to the Pauli-paramagnetic contribution, the density of states at the Fermi level, $D(\varepsilon_F)$, is calculated at 13 electrons/eV. This value is 3 times larger than the theoretical value: $D(\varepsilon_F) \approx 4$ electrons/eV for nonmagnetic $CoO_2$ according to LDA calculations [12].

It is worthwhile to compare the magnetic property of "electron-doped" $Li_{0.12}CoO_2$ to that of the parent phase, $CoO_2$. Our magnetic data (Fig. 3) demonstrate that the $\chi - T$ curve for $Li_{0.12}CoO_2$ is essentially identical to that for $CoO_2$, indicating that $Li_{0.12}CoO_2$ is also a Pauli paramagnet of the same magnetic origin as $CoO_2$. Thus, the $x = 0$ composition is *not* a singular point in the electronic phase diagram of the $A_xCoO_2$ systems. The situation is in sharp contrast to the case of high-$T_c$ superconductive copper oxides in which the parent material is a Mott insulator. The present study implies that any theoretical approaches starting from a Mott insulator may not be appropriate for discussion of the metallic state in $A_xCoO_2$.

In summary, synthesis and properties of $CoO_2$, the $x = 0$ phase of the $A_xCoO_2$ systems ($A$ = Li, Na), were reported. Single-phase bulk samples of $CoO_2$ were successfully obtained through electrochemical de-intercalation of Li from pristine $LiCoO_2$. The magnetism of $CoO_2$ is featured with a temperature-independent susceptibility with a relatively large value, being essentially identical to that of a Li-doped phase, $Li_{0.12}CoO_2$, This suggests that the $CoO_2$ phase is a Pauli-paramagnetic metal with itinerant electrons.

The present work was supported by Grants-in-Aid for Scientific Research (Contracts 16740194 and 19740201) from the Japan Society for the Promotion of Science. M.K. acknowledges financial support from the Academy of Finland (Decision 110433).

Table 1. The $\chi_0$, $C$, and $\Theta$ values for the $LiCoO_2$, $Li_{0.12}CoO_2$, and $CoO_2$ samples. These values were determined based on least-square calculations for the $\chi$ versus $T$ plots (Fig. 3).

| $x$ in $Li_xCoO_2$ | $\chi_0$ ($10^{-4}$ emu / mol Oe) | $C$ ($10^{-3}$ emu K / mol Oe) | $\Theta$ (K) |
|---|---|---|---|
| 0.0 | 5.68 | 4.01 | −3.3 |
| 0.12 | 6.00 | 5.48 | −3.1 |
| 1.0 | 1.34 | 0.89 | −1.6 |

Figure captions

Fig. 1:

Change in voltage for the $Li_xCoO_2$/Al electrochemical cell as Li is de-intercalated from $LiCoO_2$. The Li content ($x$) of $Li_xCoO_2$ is calculated based on Faraday's law.

Fig. 2:

X-ray powder diffraction patterns for (a) $CoO_2$, (b) $Li_{0.12}CoO_2$, and (c) $LiCoO_2$ samples.

Fig. 3:

Temperature dependence of magnetic susceptibility ($\chi$) for the samples of $CoO_2$ (red circles) $Li_{0.12}CoO_2$ (blue triangles), and $LiCoO_2$ (black squares).

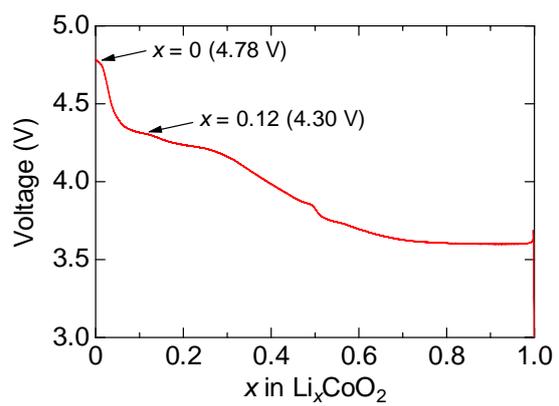

Fig. 1. Motohashi *et al.*

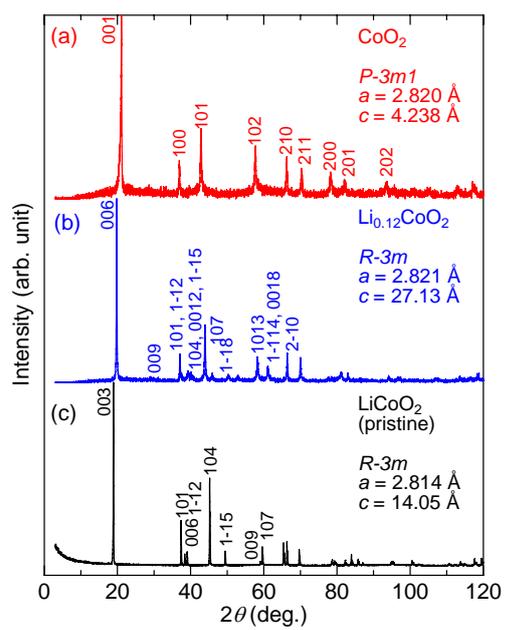

Fig. 2. Motohashi *et al.*

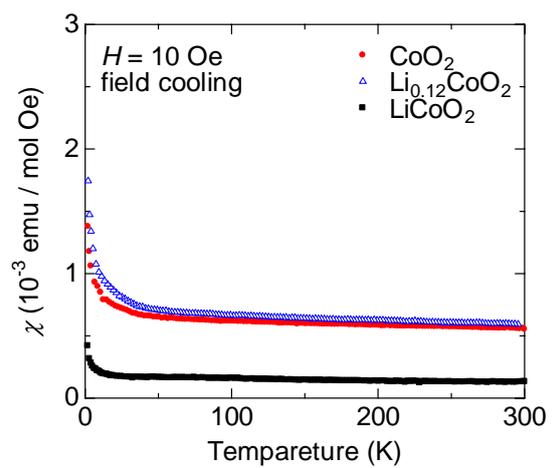

Fig. 3. Motohashi *et al.*